%% file: Karachentsev_n.tex
\documentclass[
aps,%
12pt,%
final,%
notitlepage,%
oneside,%
twocolumn,%
nobibnotes,%
nofootinbib,%
superscriptaddress,%
noshowpacs]%
{revtex4}

\usepackage[english]{babel}
\usepackage{graphicx}
\usepackage{longtable}
\input sao_cmd

\begin{document}

%

\flushbottom
%
\selectlanguage{english}
\title{Binary Galaxies in the Local Supercluster and Its Neighborhood}
\author{\firstname{I.~D.}~\surname{Karachentsev}}
\affiliation{\saoname}
\email{ikar@luna.sao.ru}
\author{\firstname{D.~I.}~\surname{Makarov}}
\affiliation{\saoname}
\affiliation{Invited Researcher Observatoire de Lyon, St-Genis Laval Cedex, France 69561}

\received{May 22, 2008}%
\revised{June 9, 2008}%

\begin{abstract}
We report a catalog of 509 pairs identified among  10403 nearby galaxies with line-of-sight
velocities
$V_{LG} < 3500$ km/s. We selected binary systems in accordance with two criteria (``bounding'' and
``temporal''),
which require  the physical pair of galaxies to have negative total energy and its components to be
located
inside the zero-velocity surface. We assume that individual galaxy masses are proportional to their
total
$K$-band luminosities, \mbox{$M = L_K \times 6M_{\odot}/L_{\odot}$.} The catalog gives the
magnitudes and
morphological
types of galaxies and also the projected (orbital) masses and pair isolation indices. The
component line-of-sight velocity differences and projected distances of the binary systems
considered
have power-law distributions with the median values of  35 km/s and 123~kpc, respectively. The
median
mass-to-$K$-band luminosity ratio is equal to $11M_{\odot}/L_{\odot}$, and its uncertainty is mostly
due to
the errors of measured velocities. Our sample of binary systems has a typical density contrast of
$\delta\rho/\rho_c\sim 500$ and a median crossing time of about 3.5~Gyr. We point out the
substantial fraction of
binary systems consisting of late-type dwarf galaxies, where the luminosities of both components are
lower than
that of the Small Magellanic Cloud. The median projected distance for 41 such pairs is only 30~kpc,
and the median
difference of their line-of-sight velocities is equal to  14 km/s which is smaller than the typical
error for
radial-velocity (30 km/s). This specific population of gas-rich dwarf binary galaxies
such as  I~Zw~18
may be
at the stage immediately before merging of its components. Such objects, which are usually lost
in flux-limited
(and not distance-limited) samples deserve a thorough study in the $HI$ radio line with high spatial
and velocity
resolution.

\end{abstract}

\maketitle

\newpage

\section{INTRODUCTION}

This is the first paper of a series devoted to the study of the visible and dark matter within the
nearby, but
sufficiently
representative volume of the Local Supercluster (LS) and its neighborhood, which is comparable to
the size of
a cosmological homogeneity cell. Over the past decade mass spectroscopic and photometric galaxy
surveys---SloanDSS, 2MASS, 2dF, and 6dF---have been performed, which reshaped and refined our concepts about
the
large-scale structure of the Universe. However, the surveys performed within certain sky areas or in
narrow strips
out
to redshifts $z \simeq$ 0.1--1.0 proved to be insufficient to allow analyses of the structure and
kinematics of small-
scale features like the Local Group, since they did not include numerous dwarf galaxies because of
their low
luminosity. For example, in the Sloan Digital Sky Survey  (SDSS)  the mean distance between galaxies
with measured
line-of-sight
velocities is equal to 9~Mpc, which exceeds the diameter of a typical cluster.

In recent years, considerable effort was focused on the study of the most nearby, so-called Local
Volume (LV)
of radius 10~Mpc, where more than 500 galaxies have been found. Most of these objects are dwarf
systems with measured
line-of-sight velocities, and about half of these galaxies already have individual distance
estimates that are
accurate to
at least 10\%, supported by observations made with the Hubble Space Telescope. The detailed 3D
pattern of the
distribution of galaxies in the LV, where the density of galaxies with measured velocities is two
orders of
magnitude higher  than the corresponding density for the SDSS, 2dF, and 6dF surveys, allowed the
structure and
kinematics of groupings to be studied on scale lengths  0.1--1~Mpc
\cite{Kar04:Karachentsev_n,Kar05:Karachentsev_n,Kar08a:Karachentsev_n}.
The contribution of the virial masses of nearby groups proved to be three-to-four times less than
the average
cosmological
density, $\Omega_m=0.27$. This inconsistency between the local and global $\Omega_m$ estimates can
be due to poor
statistics or to some specifics of our  immediate  neighborhood. An evident way for resolving this
paradox consists
in increasing the volume studied to make it include the entire Local supercluster and its immediate
neighborhood.
Tully~\cite{Tully87:Karachentsev_n,Tully88:Karachentsev_n} was the first to successfully undertake
such an analysis.
He
compiled a catalog and an atlas of nearby galaxies with line-of-sight velocities smaller than
\mbox{3000 km/s.}
Tully's catalog contains a total of 2367 galaxies located inside the volume of diameter  82~Mpc,
which is comparable
to the volume of a cosmological homogeneity cell. Tully used the hierarchical dendrogram method to
identify in this
volume a total of 179 groups, which included  69\% of all the galaxies considered. He then used the
virial masses of
these groups to infer a lower limit for the local mass density, \mbox{$\Omega_{vir}=0.08$,} which
proved to be three
times smaller than the global value of \mbox{$\Omega_m=0.27$.}

The amount of dark matter per unit luminosity of galaxies is known to increase from small groups to
rich clusters.
However, the virial regions of clusters contain only \mbox{5--10\%} of all galaxies, and about the
same number of
galaxies are associated with the unvirialized peripheral regions of these clusters. About half of
all galaxies are
members of groups like our Local Group, about one fourth of all galaxies reside in dispersed groups
(clouds), and a
total of 5--10\% of
all galaxies are located in the overall field. In such rather arbitrary and coarse partition groups
of galaxies are
the main contributors to the global mean mass density. However, the characteristic estimates of the
masses of groups
of galaxies differ by more than one order of magnitude. This circumstance emphasizes the need for
further refinement
of virial masses of groups of galaxies, which is great importance for cosmology.

Below we consider galaxies with line-of-sight velocities with respect to the Local Group
$V_{LG}<3500$ km/s. After
excluding the region of strong absorption at Galactic latitudes  $\mid b\mid <15^{\circ}$ we fixed a
total of 10403
galaxies in this volume and applied to them the criterion of identifying multiple systems.

In this paper we consider only binary galaxies, because the corresponding sample illustrates most
clearly
the specific features of the criterion employed. In our next papers we plan to present the list of
galaxy
triplets and analyze the properties of groups with four to 400 members, describe our catalog of very
isolated LSC
galaxies, and specific features of the distribution of voids. Individual
Tally--Fisher distances~\cite{Tal-Fish77:Karachentsev_n} are already available for about 1700 LSC
galaxies~\cite{Tal08:Karachentsev_n}.
We plan to use these data to analyze non-Hubble motions in the LSC in order to probe the
distribution of
dark matter on 3--10~Mpc scale lengths.

\section{CRITERIA FOR SELECTING MULTIPLE SYSTEMS OF GALAXIES}

Various algorithms have been suggested to identify groups of galaxies in a magnitude- or
distance-limited sample.
All these algorithms actually reduce to the following two main ones: percolation (the ``friend of
friend''
method) and taxonometry (construction of a hierarchical tree).

\mbox{Huchra and Geller \cite{Hu-Gel82:Karachentsev_n}} used the percolation method by
joining galaxies  into groups based on the condition that their projected mutual distances and
line-of-sight
velocity differences should be smaller than certain threshold values $R_c$ and $V_c$.
With $R_c=0.52$~Mpc and \mbox{$V_c=600$ km/s} the above authors grouped about 74\% CfA galaxies. The
resulting
groups had a typical size of  $R_H=1.1$~Mpc, line-of-sight velocity dispersion of
$\sigma_v=208$~km/s, and an average virial mass of $\lg (M_{vir}/M_{\odot})=13.5$. Many authors
applied
this method to different galaxy samples. One of the weak points of the method is freedom in the
choice
of two percolation parameters, $R_c$ and $V_c$, whose variation affects substantially the
characteristic
sizes and masses of the groups and the percentage of galaxies found to be group members.
In the percolation algorithm parameters $R_c$ and $V_c$ trace certain contrast of galaxy number
density and
overlooks many real groups in low-density regions, while clusterizing small unvirialized aggregates
in
high-density regions. Another disadvantage of the  ``friend of friend'' methods manifest itself in
the form of
the strong dependence of group parameters on the distance $D$ to the group. Numerous attempts
aimed to
reduce this dependence by introducing variable quantities $R_c(D)$ and $V_c(D)$ resulted in
subjectively
arbitrary choices. The most recent application of the percolation method to  2MASS
\cite{Crook07:Karachentsev_n}
galaxies
yielded  1258 groups and 1710 pairs of galaxies for a relative density contrast of
\mbox{$\delta\rho/\rho=80$.}
In general, members of groups and pairs make up for a total of  36\% and 17\% of the entire
sample.
Groups with $n\geq 5$ elements have a projected radius of about 1~Mpc, line-of-sight velocity
dispersion
on the order of~\mbox{200~km/s}, and an average virial mass of
\mbox{$\lg(M_{vir}/M_{\odot})\sim13.5$.}
At the depth of the  2MASS sample \mbox{($D_{max}=140$~Mpc)} the contribution of virial
masses of
groups identified using the percolation method is equal to only \mbox{$\Omega_m$=0.10--0.13.} An examination
of the
list of these groups gives a rise to numerous questions. In particular, we do not understand why
Eridanus+Fornax~I is the most massive cluster complex instead of Virgo, which we know as the center of the Local Supercluster.

Tully~\cite{Tully87:Karachentsev_n,Tully88:Karachentsev_n} and Vennik  \cite{Ven84:Karachentsev_n}
used another, `taxo\-nometric'', method to group galaxies into pairs in accordance with the maximal ratio of
luminosity to cubed
mutual distance ($L_{ik}/R_{ik}^3$). The resulting pair was substituted by a  ``particle''
with the luminosity equal to the total of the galaxies and the search for maximal ($L_{ik}/R_{ik}^3$) was
repeated.
The process ended by the construction of a single hierarchical ``tree'' with branches containing
the entire
galaxy sample considered. Cutting the tree branches at a certain level of the contrast of volume
density
yielded a set of branches (groups) whose sizes and virial masses depended on the adopted density
(luminosity)
contrast. Tully~\cite{Tully87:Karachentsev_n} used the method of dendrograms to infer a
characteristic group radius of  0.32~Mpc and the average line-of-sight velocity dispersion of
\mbox{$M_{vir}/L_B=94 M_{\odot}/L_{\odot}$,} which proved to be substantially lower than the average
ratio
for groups of the Huchra--Geller~\cite{Hu-Gel82:Karachentsev_n} list.
Practical applications of both the percolation and the dendrogram methods ignored individual
properties of galaxies
by viewing them as indistinguishable particles. It is evident that the same threshold values
$R_c$ and $V_c$ would be sufficient (and even redundant) for clustering a pair of dwarf galaxies
and, at the same time,
evidently insufficient to bind a pair of giant galaxies. Such inadequacy of the criterion distorts
the estimates of virial masses.

\section{CLUSTERIZATION ALGORITHM}

Galaxies can be grouped into small systems with their individual properties taken into account
by viewing
two arbitrary galaxies as a virtual bounded pair~\cite{Kar94:Karachentsev_n}.
We proceed from this evident premise and require that the difference $V_{ik}$ of the space
velocities
of galaxies in physical pair and their mutual space distance $R_{ik}$  obey the condition of
negative
total energy
$$
   \frac{V_{ik}^2R_{ik}}{2GM_{ik}}<1, \eqno(1)
$$
where $M_{ik}$ is the total mass of the pair and $G$ is the gravitational constant. We correct the
squared
velocity difference of the pair $V_{ik}^2$ for velocity measurement errors. However,
observations
give us only the line-of-sight projection of velocity $ V_{ik}$ and the sky-plane projection of
$R_{ik}$.
Therefore condition~(1) must be supplemented by an additional constraint onto the maximum distance
between
the components for fixed mass $M_{ik}$. The condition that the components of the pair are located
inside
the  ``zero-velocity'' sphere
\cite{San86:Karachentsev_n}
has the following form
$$
  \frac{\pi H^2R_{ik}^3}{8GM_{ik}} <1, \eqno(2)
$$
where $H$ is the Hubble constant. Note that both conditions  (1) and (2) are conservative with
respect
to projection factors, i.e., use of projected mutual velocities and distances in formulae (1)
and (2)
instead of the space velocities and distances does not exclude true (physical) pairs. However, these
conditions
do not prevent false (optical) pairs from getting into the catalog.

Our algorithm is actually a variant of the percolation method. We first identify all pairs
satisfying conditions (1)
and (2) and then group all pairs having a common component into a single entity. Finally, if we find
a galaxy
to be a satellite of several more massive galaxies, we link it to the most massive neighbor. In
particular,
a group may be a subgroup inside a more massive structure. In this sense, our algorithm
combines the advantages of both the  ``friend of friend'' method and hierarchical approach.

We determine the masses of galaxies from their $К$-band IR luminosity assuming that all galaxies
have the same
``mass--luminosity'' ratio:
$$
  M/L_K={\kappa} (M_{\odot}/L_{\odot}), \eqno(3)
$$
where we set $\kappa$ equal to  6.
In our algorithm \mbox{$\kappa=6$} is actually the only more or less arbitrary quantity. We chose it
based on
the following assumptions. According to the data of \mbox{Bell et al.
\cite{Bell03:Karachentsev_n},}
the average cosmic mass-to-\mbox{$K$-band-luminosity} ratio is equal to
\mbox{$0.95\pm0.27M_{\odot}/L_{\odot}$,} which agrees well with the results of the computations of
\mbox{Fukijita et al. \cite{Fuk96:Karachentsev_n}.}
According to the data compiled by Karachentsev and Kut'kin~\cite{Kar+Kut05:Karachentsev_n}, for
galaxies of
the Local Volume the  average ratio of the mass inside standard radius $R_{25}$ to the integrated
$K$-band luminosity is equal to \mbox{$\langle M_{25}/L_K\rangle =1.3\pm0.2$} in solar units, and
this ratio remains almost constant and varies from  \mbox{$1.1\pm0.2$} for giant galaxies with
the mean luminosity of \mbox{$1.1\times 10^{11}L_{\odot}$} to \mbox{$1.5\pm0.2$} for dwarf galaxies
with the mean  $K$-band luminosity of \mbox{$1.1\times 10^7L_{\odot}$} (the slight increase of
$\langle M_{25}/L_K\rangle $  toward dwarf galaxies is evidently due to higher content of the
gaseous
component). Flat rotation curves observed for most of galaxies indicate the dominating
role of the dark halo beyond the standard radius $R_{25}$. The most extended rotation
curves
inferred from the data of the  \mbox{21-cm line} emission reach out to \mbox{$R_{max}=
(3$--$6)R_{25}$
\cite{Bro92:Karachentsev_n,Swa02:Karachentsev_n,Beg05:Karachentsev_n}.}
These  $R_{max}$ values correspond to the global ratio of \mbox{$M/L_K\simeq 6
M_{\odot}/L_{\odot}$}
\mbox{used in formula (3).} Note that we ``trained''  clusterization algorithm \mbox{(1--3)} by
applying
it  to the detailed 3D distribution of galaxies in the Local Volume.
Choosing dimensionless parameter $\kappa$ values in the  $\kappa<4$ interval drastically
reduces the relative number of clusterized galaxies, whereas adopting
\mbox{$\kappa>10$} results in grouping of galaxies into extended and evidently nonvirialized
aggregates.
With the value $\kappa=6$ adopted in this paper the dwarf companions in  known nearby groups are
usually
located inside the zero-velocity spheres around massive galaxies of the corresponding groups. Note
also
that the average virial mass-to-luminosity ratio \mbox{$\langle M_{vir}/L_K\rangle =
17.5\pm3.6 M_{\odot}/L_{\odot}$} for eight groups of the Local Volume:
Local Group, M81, CenA, M83, IC342, Maffei, LeoI и  NGC6946 (whose mean luminosity is
$\langle L_K\rangle = 1.3\times 10^{11}L_{\odot}$) is almost equal to the typical
$M_{vir}/L_K$ ratio that we inferred for small groups of the same luminosity inside the entire
volume
of the Local Superclusrter. \mbox{Guzik and Seljak \cite{Gu+Se02:Karachentsev_n}} found a similar
total mass-to-$K$-band luminosity ratio---\mbox{($17.0\pm2.9 M_{\odot}/L_{\odot}$)}---for
small groups and field galaxies with \mbox{$\langle L_K\rangle \sim0.8\times 10^{11} L_{\odot}$} by
analyzing effects of gravitational lensing, and this fact demonstrates the good agreement
of the galaxy group mass estimates obtained by two independent methods.

\section{INITIAL OBSERVATIONAL DATA}

Our main sources of the data on line-of-sight velocities, apparent magnitudes, morphological types,
and
other parameters of galaxies are the  HyperLEDA~\footnote{{\tt http://leda.univ-lyon1.fr}}
\cite{Pat03:Karachentsev_n}
and NED\footnote{{\tt http://nedwww.ipac.caltech.edu/}} databases.
Note that these databases contain a substantial number of objects with erroneous line-of-sight
velocities
adopted from automatic sky surveys like 6dF. Cases of confusion of coordinates and velocities are
rather
common for galaxies located closely to each other on the celestial sphere. The apparent magnitudes and
line-of-sight velocities in the  SDSS survey often correspond to individual knots and associations
in
bright galaxies. These effects are very important for selecting of true close pairs of galaxies.
We took these effects into account and made necessary corrections, where
possible. This
proved to be most
time-consuming part of our work. Because the databases are permanently updated with new (and
sometimes erroneous)
data, it is necessary to repeat the correction of information. That is why we fixed the sample of initial
data as
it was in mid-2006 (i.e., June, 2006).

We independently found optical identifications for many $HI$ sources of the  HIPASS survey by
refining their
coordinates and determining the apparent magnitudes and morphological types of galaxies
\cite{Kar08b:Karachentsev_n}.
We examined many dwarf galaxies, especially those with low surface brightness, on the DSS digital
images
in order to find their principal parameters. The typical error of our visual estimates of
galaxy
magnitudes is about $0.5^m$, and the average error of the inferred type is about $\pm2$ in the
numeric scale
employed by de Vaucoulers in the  \mbox{RC2 catalog \cite{Vau76:Karachentsev_n}.}
The  best indicator of the baryonic mass of a galaxy is known to be its infrared magnitude, which
depends only slightly on the amount of dust and the number of young stellar complexes. Thus we
adopted the longest-wavelength, $K$-band ($\lambda=2.16~\mu$m) part of the all-sky \mbox{2MASS
\cite{Jar00:Karachentsev_n,Jar03:Karachentsev_n}}
as our main source of photometry. We converted the estimates of galaxy magnitudes in other
optical
($B,V,R,I)$ and near-infrared bands ($J,H$) into the $K$-band magnitude using syntehtic galaxy
colors of
Bizzoni~\cite{Bi05:Karachentsev_n} and Fukujita et
al.~\cite{Fu95:Karachentsev_n}.
The greatest amount of photometric data is available in the $B$-band. We use the following relations
between
the $B-K$ color excess and morphological type discussed by Jarett~et~al.~\cite{Jar03:Karachentsev_n}
and \mbox{Karachentsev and Kut'kin~\cite{Kar+Kut05:Karachentsev_n}}
:

$\langle B-K\rangle =+4.10$, for galaxies of types $T\leq2$ (i.e., E, S0, and Sa),
which are dominated by bulges,

$\langle B-K\rangle =+2.35$
for late-type galaxies $T\geq9$ (i.e., Sm, Im, Irr), and

$\langle B-K\rangle =4.60-0.25\times T$  for intermediate-type ($T=3$--8) objects.

Note that due to the short exposures the 2MASS survey proved to be insensitive to low-surface
brightness
and blue galaxies. For about one thousand dwarf and spheroidal galaxies recently discovered by
Karachentseva
et al. \cite{Kara98:Karachentsev_n,Kara00:Karachentsev_n}
in the volume of the Local Supercluster, only eye-estimated  $B$-band magnitudes are available, which
we converted
into the $K$-band magnitudes using the method described above. Despite the lack of good photometry
for these
objects, gas-rich  dIr galaxies have accurate 21-cm line radial velocities and they are important
``test particles''. Due to the low luminosities of dwarf galaxies, large errors of estimated
magnitudes have virtually no effect on the results of clusterization performed using our
algorithm.

We collected all the line-of-sight velocity measurements available in the  HyperLEDA and NED
databases for
galaxies in the Local Supercluster and its neighborhood. We excluded unreliable and inaccurate
measurements, which velocity-measurement error exceeds 75 km/s. In automatic surveys  (SDSS,
2dF, and 6dF)
we also excluded the measurements with velocities smaller than 600~km/s, because they mostly represent
Milky-Way stars projected onto distant galaxies. If several line-of-sight velocity measurements
were available
for a galaxy, we chose the median one, the velocity error was estimated as the dispersion of all
measurements with
the exclusion of the outliers.

Our initial sample cleaned from unreliable and doubtful cases contained a total of  10403 galaxies
with the line-of-sight velocities  $V_{LG}<3500$ km/s, located at the Galactic latitudes
\mbox{$\mid b\mid >15\degr $.} For all these galaxies the apparent magnitudes and morphological
types have been found. To prevent the distortion of the clusterization process at the
boundary of the volume considered, we also use the data on the galaxies located in the boundary
regions with \mbox{$10\degr <\mid b\mid <15\degr$} and with \mbox{$3500<V_{LG}<4000$ km/s,} because they
may contain some of the members of galaxy groups with high virial velocities. Grouping criteria
(1--3) allowed us to identify within the Local Supercluster volume a total of  1018 galaxies belonging to binary systems.

\section{THE CATALOG OF 509 NEARBY GALAXY PAIRS}

The Table contains the binary galaxies that we identified using our criteria. The first column gives
the number of the pair in the catalog. The second column gives the name of the galaxy or its number
in the well-known NGC, IC, UGC, CGCG, ESO, PGC, and DDO catalogs or in the 2MASS, 6dF, APMUKS,
SDSS, IRAS, and other sky surveys as given in the NED. Note that for practical reasons we omit the
coordinate part of the galaxy name used in surveys. The third column gives the galaxy
coordinates for the epoch of  2000.0. The fainter component of the pair follows the
primary component and the pairs are sorted in right ascension. Columns (4) and (5)  give the
line-of-sight velocity of the galaxy (in km/s) with respect to the centroid of the Local group and
its standard error, respectively. Columns (6) and (7) give the coded morphological type and
the apparent $K$-band magnitude corrected for Galactic extinction according to
Schlegel et al.~\cite{Schle98:Karachentsev_n}, respectively.
Column~(8) gives the projected separation between the galaxies in kpc. Column~(9) gives the
logarithm of the total $K$-band luminosity of the pair. Columns~(10) and (11) give the orbital
mass-to-luminosity ratio with and without velocity measurement correction, respectively. We
compute the mass by the following formula~\cite{He85:Karachentsev_n}
$$
M_{p}=\frac{16}{G\pi} \Delta V^2R_{\bot} \eqno(4)
$$
using the component line-of-sight velocity difference and projected linear separation. We
compute the distance to the pair from the average line-of-sight velocity with respect to
the centroiod of the Local group with the adopted Hubble-constant value of  \mbox{$H=73$ km/s/Mpc.}
Column~(12) gives the logarithm of the smallest of the values given by criteria (1) and (2) for the
given pair with respect to the surrounding galaxies. The higher is this quantity, the greater is
the degree of isolation of the considered pair. A close-to-zero value implies that the pair
is at the threshold of the formation of a bigger structure (``capturing'' of a new member or
``joining'' other groups). We compute the total luminosity of the pair assuming that the absolute
magnitude of the Sun is \mbox{$K_s=3.28^m$ \cite{BM1998:Karachentsev_n}.}
We omit the negative unbiased estimated of the orbital mass for the pairs with line-of-sight
velocity differences smaller than the corresponding measurement errors.

\section{DISTRIBUTIONS OF THE PRINCIPAL PARAMETERS OF THE PAIRS}
The fraction of galaxies that are members of binary systems in the Local Supercluster and its
neighborhood is about 10\%, which is somewhat lower than the corresponding values \mbox{12--17\%,}
according to the data of Huchra and Geller~\cite{Hu-Gel82:Karachentsev_n}, Crook et al.~\cite{Crook07:Karachentsev_n},
Magtesyan~\cite{Mag88:Karachentsev_n}, and Gourgoulhon et
al.~\cite{Gou92:Karachentsev_n}.
Figure~1 shows the distribution of the mean line-of-sight velocity of the pairs. The median of this
distribution is equal to~\mbox{2389~km/s.} Figure~2 shows the map of the distribution of pair
centers on the sky
in equatorial coordinates. The region of strong Galactic absorption is shown by gray color.
It is obvious from this map and from 3D distribution (Fig.~3) that the pairs do not reveal clear large
scale structure.
However, on short scale lengths in the vicinity of
1~Mpc, the pairs exhibit excess of mutual association compared to uniform random
distribution. The pairs 21+22 and 194+195 are examples of such close associations.
\begin{figure*}[tbp]
\includegraphics[scale=0.9,angle=-90]{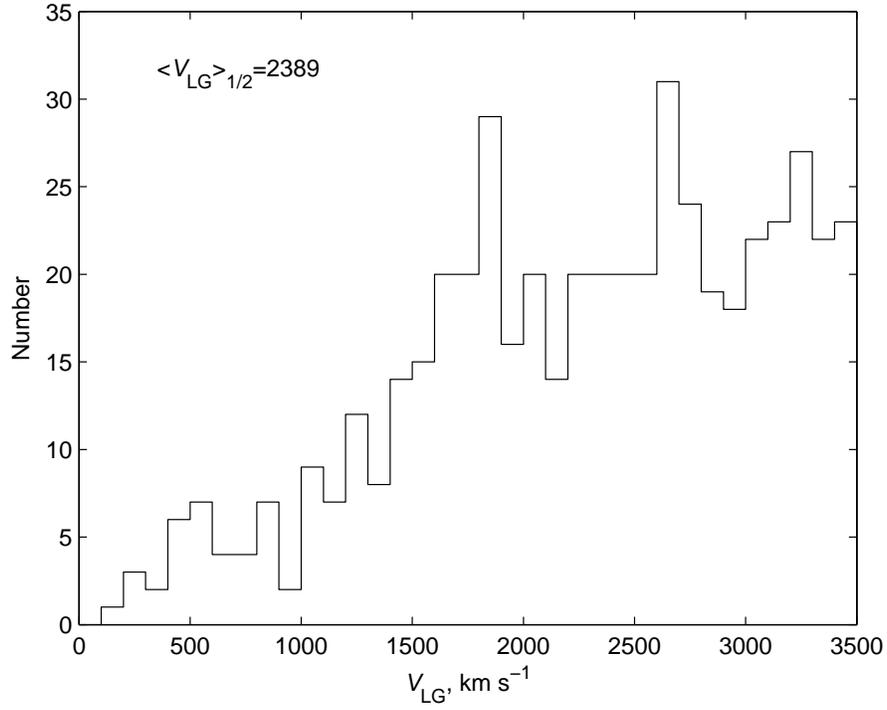}
\caption{The distribution of the average line-of-sight velocities of pairs with respect to the Local
Group.}
\end{figure*}

\begin{figure*}[tbp]
\includegraphics[scale=1.0,angle=-90]{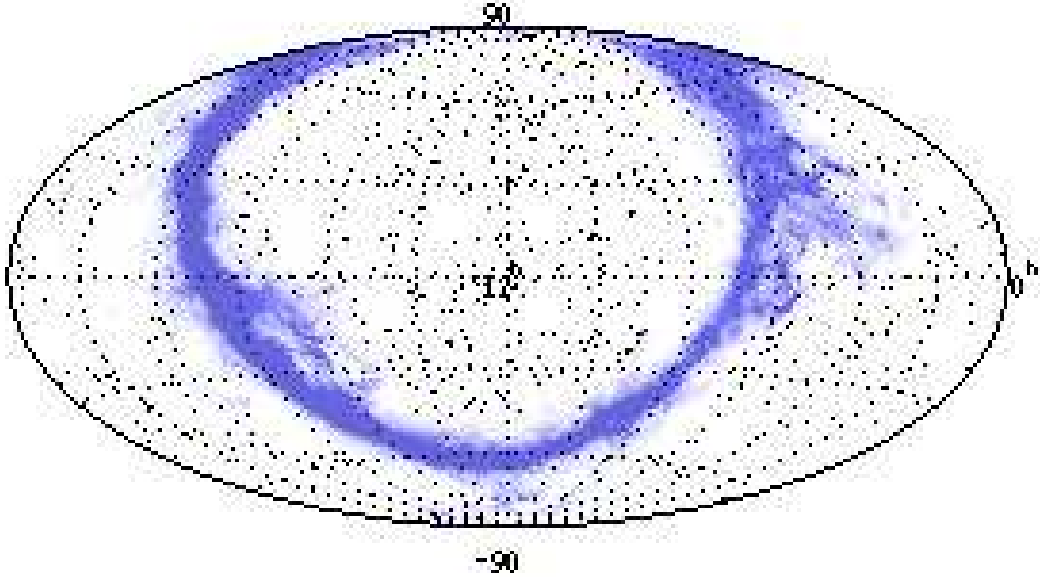}
\caption{The celestial distribution of pairs in the Local Supercluster in equatorial coordinates. The
clumpy gray region
indicates the domain of strong Galactic absorption.}
\end{figure*}

Figure~4 shows the distribution of the line-of-sight velocity difference between the satellite and
the
primary component of the pair. The distribution has quite symmetric shape with a mean difference
of
\mbox{$-1\pm 3$ km/s,} which indicates that our sample exhibits no excess of positive velocities
discussed
by Arp~\cite{Arp82:Karachentsev_n} in cases of the companions of M31 and other galaxies. The
root-mean-squared
velocity difference for the components of  509 pairs is equal to $\sigma_v=62$ km/s. Note that in the
vicinity of zero
the distribution is much sharper than a Gaussian. Note also that for about 60\% of all pairs the
velocity
difference is smaller than its error of measurement. For such pairs unbiased estimates of orbital masses
are negative and we do not give them in Column~11 of the table. Taking into account broadening of
distribution by errors,
the true distribution in \mbox{Fig.~4} must have an even sharper peak at zero velocity difference.
Note the measurement errors of velocity must be reduced several times to improve the estimation of mean
mass of pair.
\begin{figure*}[tbp]
\includegraphics[scale=0.9,angle=-90]{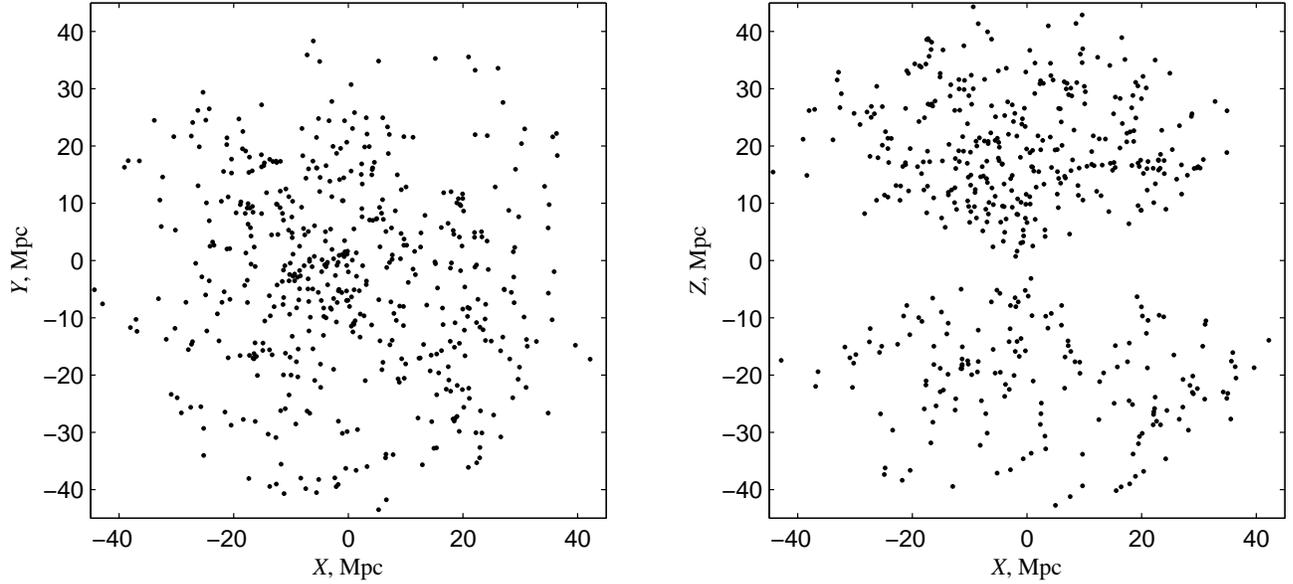}
\caption{Projected distribution of pairs in the Local Supercluster in Cartesian coordinates.}
\end{figure*}

\begin{figure*}[tbp]
\includegraphics[scale=0.9,angle=-90]{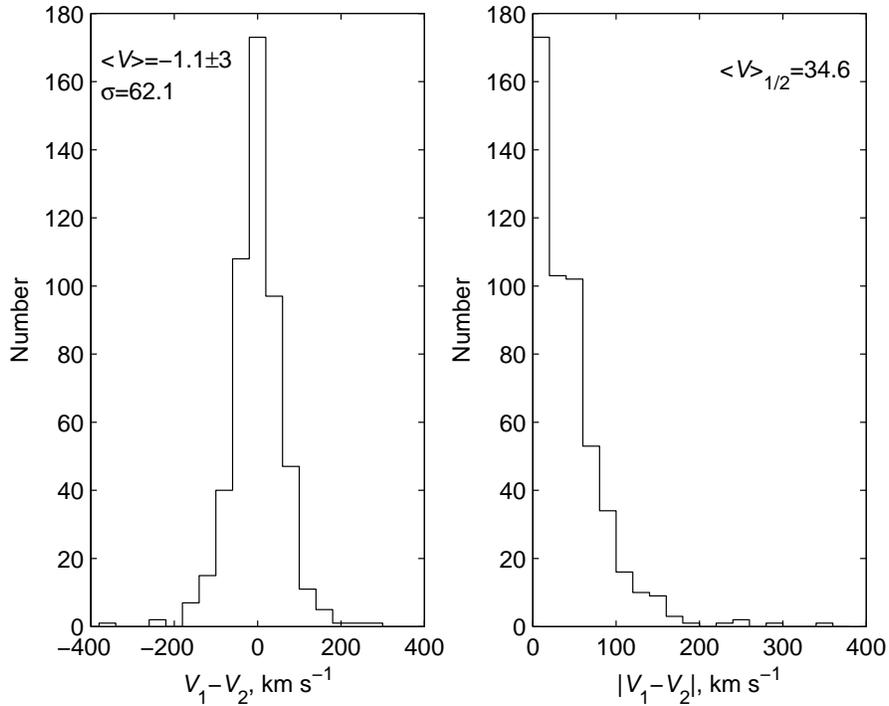}
\caption{The distribution of the companion minus primary line-of-sight velocity difference.}
\end{figure*}

Figure~5 shows the distribution of the projected linear separation between the components in  509
pairs.
The median and mean separations are equal to  123 and 177~Mpc, respectively, and the separation in
the
widest pairs reaches 1~Mpc. In general, the population of binary galaxies outlines fairly well
the linear size of a typical dark halo in the $\Lambda$CDM model. The distribution $N(R_{\bot})$ can
be
fitted by a power law with exponent an $\alpha=-1.1$. Figure~6 shows the two-dimensional distribution
of
the line-of-sight velocity difference and projected separation between the components of the pairs
in
logarithmic scale. Despite the strong effect of projection factors, pairs show a tendency toward a
decrease
of $\Delta V$ with decreasing separation between the components, however, the slope of the
regression line (the
dashed line) is very close to zero and the line differs significantly from the Keplerian law (the
solid line).
This envelope corresponds to values $\Delta V=121$ km/s and $R=750$~kpc, which is close to Andromeda and
Milky Way as wide pair of galaxies.
Note that normalization of  $\Delta V$ by the total luminosity of the pair makes the slope of the
regression line closer to the Keplerian value.

\begin{figure*}[tbp]
\includegraphics[scale=0.8,angle=-90]{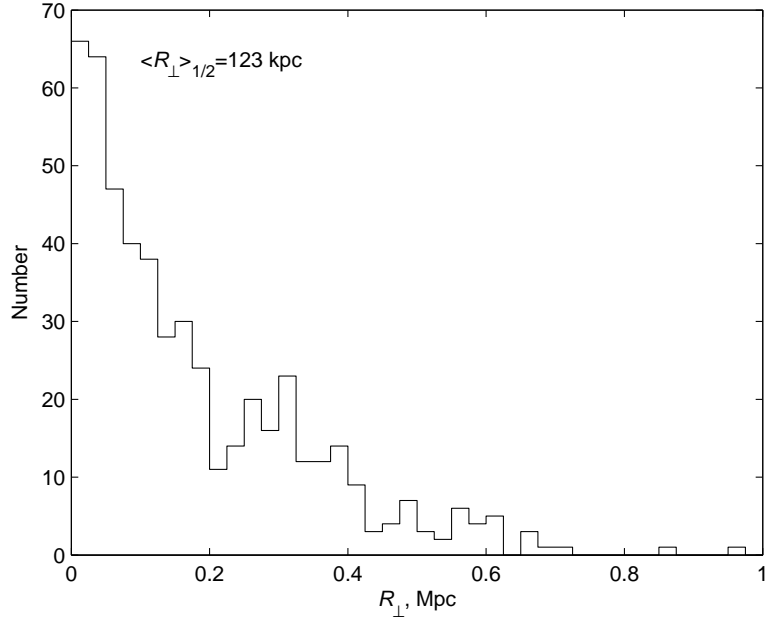}
\caption{The distribution of the projected separation between the components in the pairs of the Local
Supercluster.}
\end{figure*}

\begin{figure*}[tbp]
\includegraphics[scale=1.0,angle=-90]{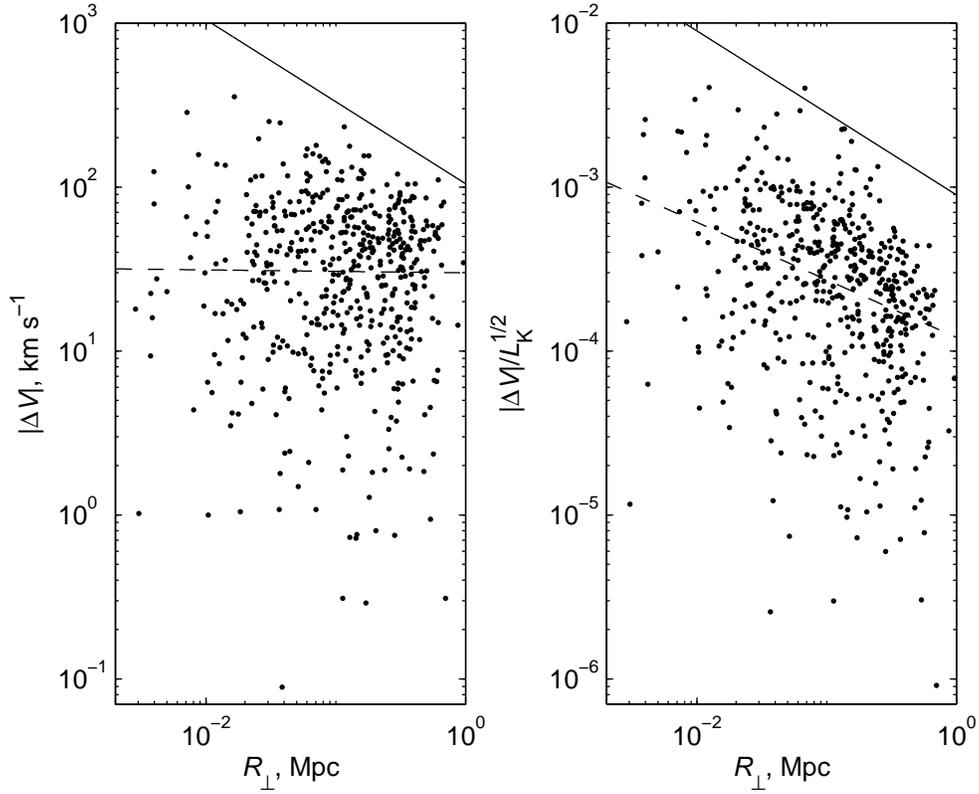}
\caption{The distribution of the line-of-sight velocity difference and projected separation for the
pairs of the
Local Supercluster.}
\end{figure*}

\begin{figure*}[tbp]
\includegraphics[scale=0.9,angle=-90]{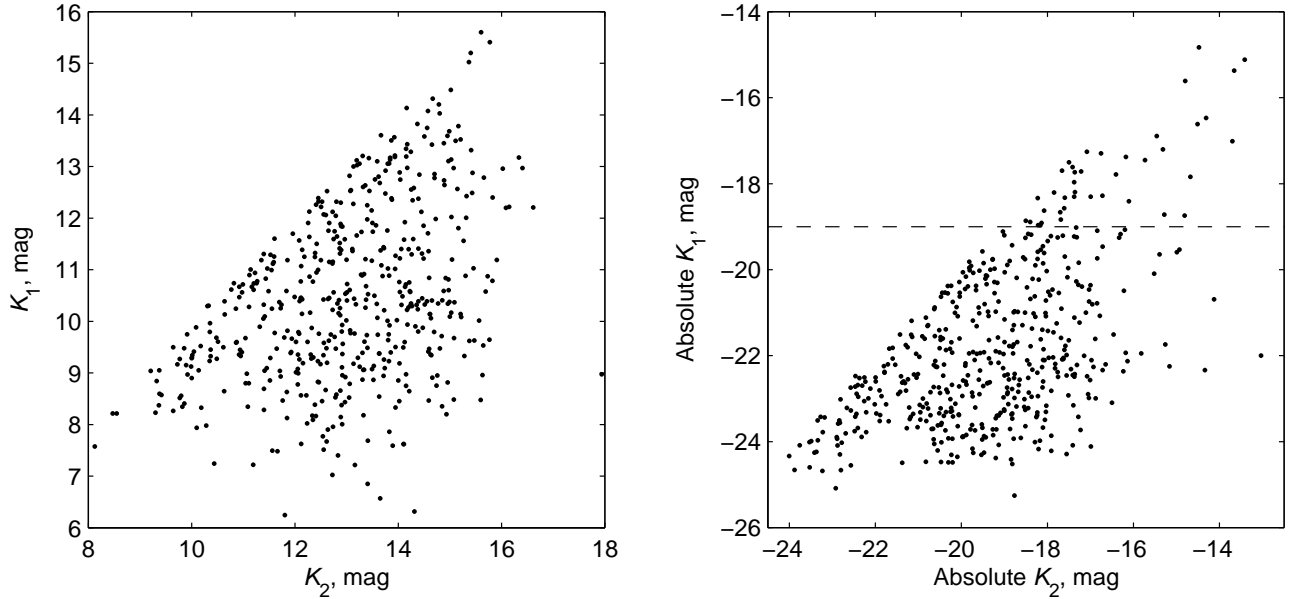}
\caption{Distribution of the apparent (left) and absolute (right) $K$-band magnitudes of the bright
(1) and faint (2)
components in the Local Supercluster.}
\end{figure*}

The left- and right-hand panels in Figure~7 show the distributions of the visual (left) and
absolute (right) $K$-band magnitudes of the bright (1) and faint (2) components of the pairs. Both
distributions fill
more or less uniformly a wide sector of possible values over 10-magnitude range. About
40\% of all pairs components differ in luminosity less than a 10 times.
At the same time, there are galaxies, e.g., NGC 3044, NGC 2683, and NGC 3621, with dwarf
companions that
are \mbox{7--10} magnitudes fainter than the primary component. The special interest are drawn by the
pairs located in the
top right corner of the right-hand panel in Fig.~7. The dashed line in this panel corresponds to
$M_K=-19.0$ (the luminosity of the  SMC). Pairs that are located above this line have both components
that are
dwarf galaxies. There are total of 41 pairs like UGC~5272 and its companion (pair No.~159), which
are located in the
domain of  lower-than-SMC luminosity. The pair of metal-poor BCD dwarfs  SAO0822+3545 and
SDSSJ0825+3532
(pair No.~113) studied by Cnehgalur et al.~\cite{Che06:Karachentsev_n} is another example.
Almost all these binary dwarfs are gas rich and contain young blue stellar population. The average
component
line-of-sight velocity difference for these pairs is equal to only 25~km/s, and the median and mean
projected
separations are equal to 30 and 42~kpc, respectively. These pairs of blue dwarfs with
considerable
reserves of gas and active star formation  may be a sort of small multiple systems at a stage close
to component
merger. On the other hand, \mbox{Tully et al.
\cite{Tal06:Karachentsev_n}} pointed out the presence in the Local Volume of groups
which exclusively consist of dwarf galaxies.
Note that old percolation criteria proved to be insensitive to multiple dwarf systems. They were
discovered
as a result of, among other things, in-depth analysis of the population of the Local Volume and of
use a more refined algorithm for group searching. It is evident that $HI$-observations
with
high angular resolution would be a promising method for analyzing the kinematics and evolutionary
status of dwarf pairs pairs and groups.
\begin{figure*}[tbp]
\includegraphics[scale=0.9,angle=-90]{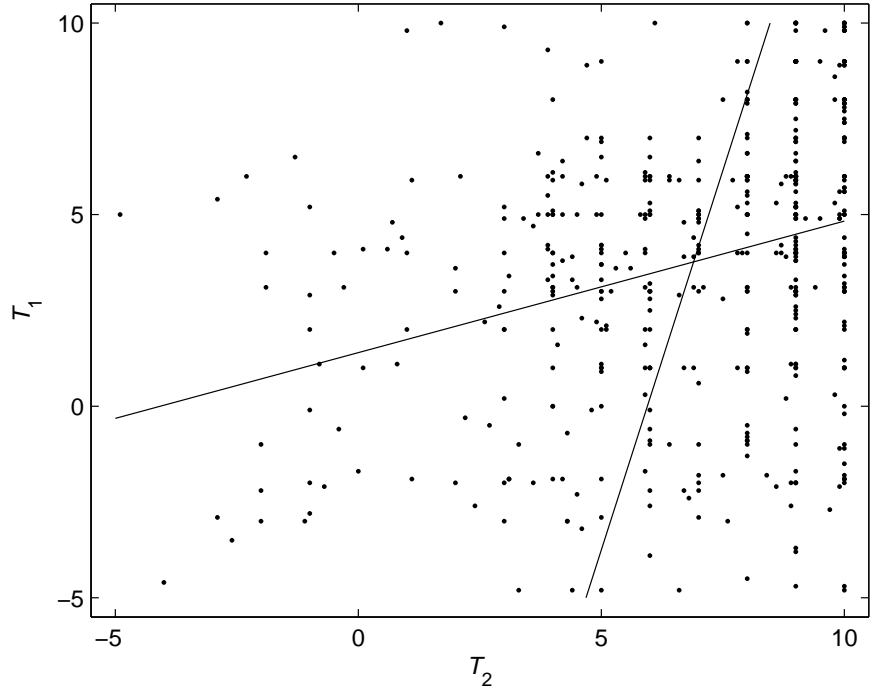}
\caption{Distribution of the types of the bright (1) and faint components. The lines show the direct
and
invrse regression}
\end{figure*}

Figure~8 shows the distribution of the morphological types of the bright (1) and faint (2)
components of 509 pairs. On the average, the primary galaxy have an earlier type
($\langle T_1\rangle =3.8$)
than its companion  (\mbox{$\langle T_2\rangle =6.9$)}. This fact is
easy to explain by the well-known correlation between the luminosity and morphological type of
galaxies. The lines of direct and reverse regression in Fig.~8 point out a weak correlation
between the morphological types of the components, which also can be due to the luminosity
effect.
\begin{figure*}[tbp]
\includegraphics[scale=0.8,angle=-90]{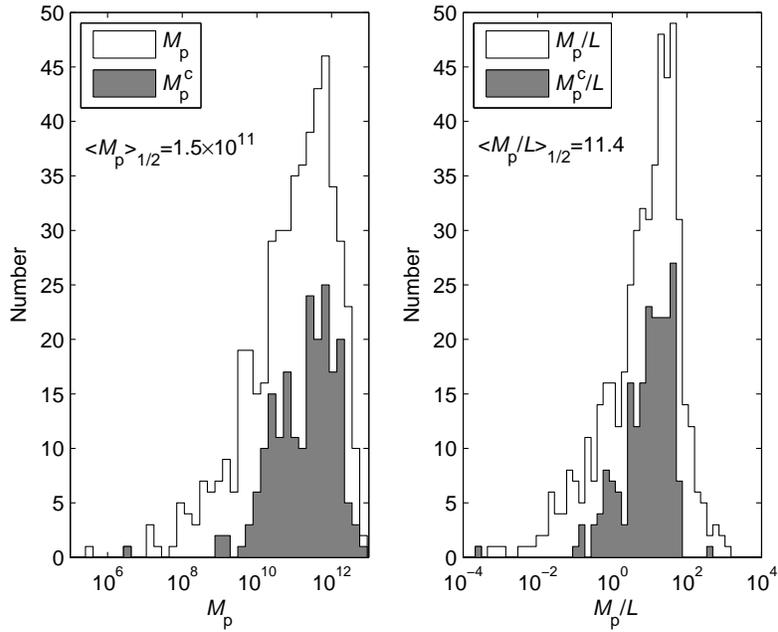}
\caption{Distribution of estimated orbital masses and orbital mass-to-luminosity ratios
for the pairs of the  Local Supercluster.}
\end{figure*}

Figure~9 shows the distribution of estimated orbital masses and orbital mass-to-luminosity ratios
for galaxy pairs inferred in accordance with formula (4). The median mass of the pairs is
\mbox{$1.5\times 10^{11}M_{\odot}$,}  and the median mass-to-luminosity ratio is $11.3
M_{\odot}/L_{\odot}$, which is almost twice the $\kappa=6$ value that we adopt for individual
galaxies. However, these mass estimates are statistically biased. We already pointed out above
that the line-of-sight velocity difference is smaller than its standard error for more than half
of all pairs. To obtain an unbiased mass estimate, we must substitute $(V^2_{12} - \sigma^2_{1}-
\sigma^2_{2} $) for $V^2_{12}$ in formula~(4). In this case mass estimations become negative for
60\% of the pairs and the median mass-to-luminosity ratio also becomes negative ($-3
M_{\odot}/L_{\odot}$). The distributions of  $M$ and $M/L$ for the pairs corrected to for
velocity measurement errors are shown in gray, but they represent only the domain of positive
masses. All these data illustrate the fact that the accuracy of line-of-sight
velocities measured in modern optical spectroscopic galaxy surveys (2dF, SDSS, 6dF), which is
about 50~km/s, is
clearly insufficient to compute a bona fide average mass of galaxy pairs. Obviously
the special observations of binary galaxies are needed to reduce the velocity errors down
to 5--10~km/s.
\begin{figure*}[tbp]
\includegraphics[scale=0.8]{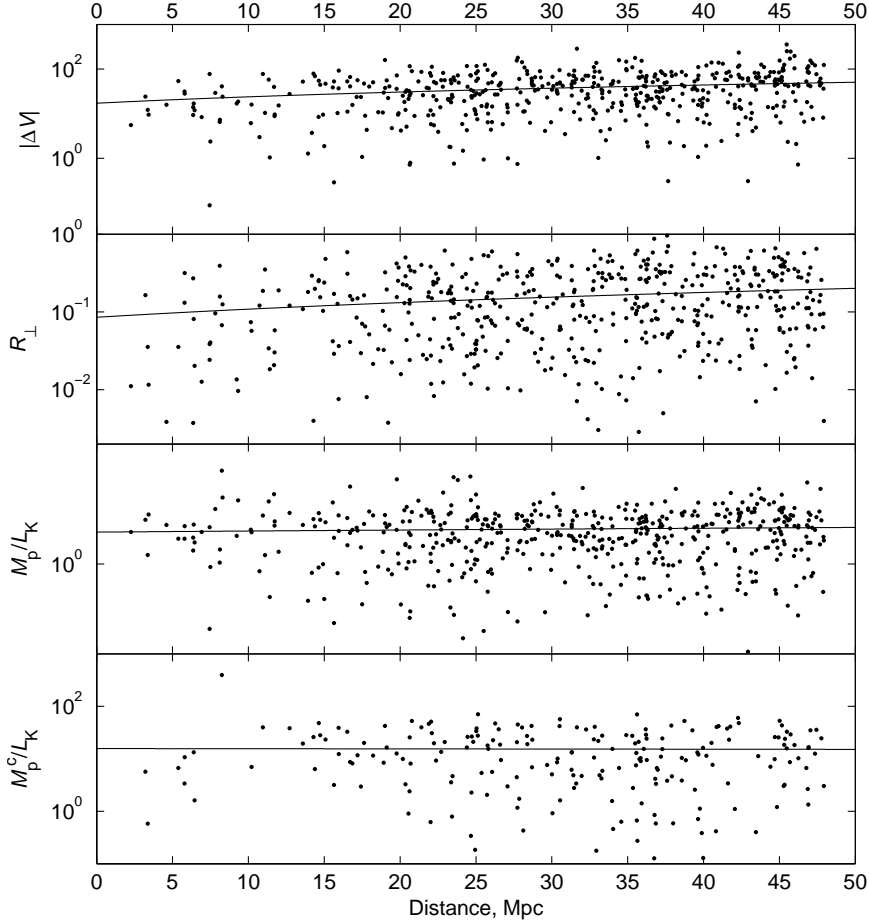}
\caption{Characteristic diagrams showing how the properties of identified galaxy pairs
vary with distance.}
\end{figure*}

\begin{figure*}[tbp]
\includegraphics[scale=0.8,angle=-90]{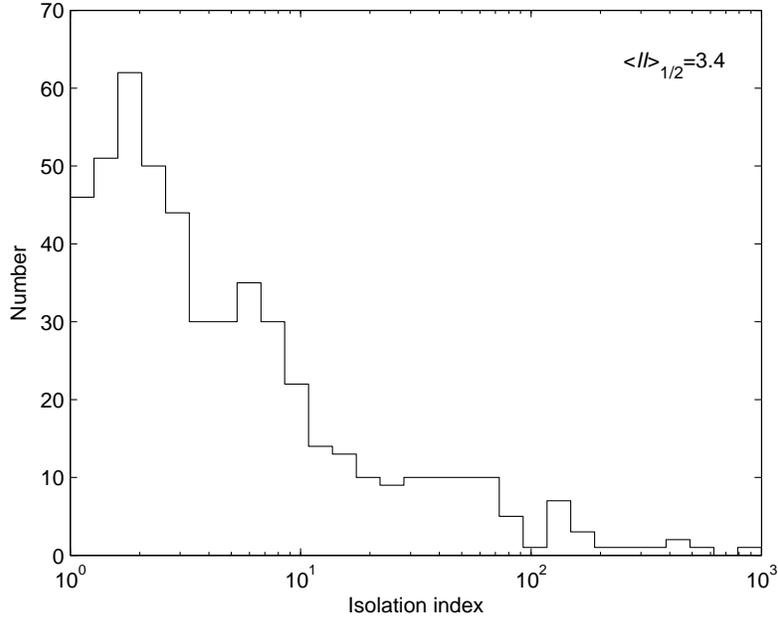}
\caption{Distribution of the index of isolation of the pairs of the Local Supercluster
with respect to neighboring galaxies.}
\end{figure*}

Figure~10 shows how the line-of-sight velocity difference, projected separation between the
components, and the orbital mass-to-luminosity ratio of binary galaxies vary with heliocentric
distance. It is evident that in our sample the fraction of pairs consisting of two dwarf components
decreases appreciably toward the outskirts of the volume considered. However, the average
dynamical parameters of the ensemble of galaxy pairs vary little with distance. This fact again
demonstrates the advantage of our criterion, which takes individual properties of galaxies into
account.

\section{COMPARISON WITH OTHER \protect\linebreak SAMPLES OF GALAXY PAIRS}

The most detailed study of binary galaxies was performed by Karachentsev~\cite{Kar87:Karachentsev_n}.
He compiled a catalog of 603 isolated Northern-sky pairs with components brighter
than $B=15.7^m$ (they are denoted as KPG in the NED database). These pairs were identified on the
condition that they are isolated with respect to nearby projected galaxies without invoking
line-of-sight velocity data. Later Reduzzi and Rampazzo~\cite{Re+Ra95:Karachentsev_n}
used the same criterion to identify a total of 409 pairs in the Southern sky (these pairs
are denoted as RR in the NED database). The isolation condition favored the identification of
closer binary systems with brighter components. The characteristic depth of the KPG
sample is 6350 km/s, which is substantially greater than the Local Supercluster sample
considered in current paper. After taking into account various selection effects in the catalog of isolated pairs we
estimate the fraction of galaxies in binary systems to be $12\pm2$\%. The KPG pairs with their
order-of-magnitude higher luminosities \mbox{(about $1.3\times 10^{11}L_{\odot}$}
if transformed into the  $K$-band values) have large line-of-sight velocity differences,
\mbox{$\langle V_{12}\rangle =120$ km/s.} With the average component separation of only $\langle
R_{\bot}\rangle  =40$~kpc, isolated pairs have a moderate orbital mass-to-luminosity
ratio of $\langle M/L_B\rangle =7.8 M_{\odot}/L_{\odot}$ and almost do not show evidence for dark
halo on these scale lengths.

Our list contains a total of 16 pairs located within the rather thoroughly studied Local Volume. In
cases where individual distances have been measured for components of nearby pairs these
distances confirm the relative closeness of the pair members. However, the distances  of
the NGC 4449 and UGC 7577 galaxies (pair No.~281) (\mbox{4.21} and \mbox{2.54~Mpc}, respectively)
indicate that these galaxies are just accidentally located along the same line of sight.
Despite the small  line-of-sight velocity difference (\mbox{252 km/s} and \mbox{240 km/s}) and rather isolated
location ($II=2.3$) these members of the CVnI cloud cannot be viewed as an isolated physical pair.

Among the isolated pairs of the KPG catalog there are several known nearby pairs, in particular,
NGC 5194 + NGC 5195 (M51) and NGC 672 + IC 1727. Although our criterion (1--3) clusterizes these
galaxies, it nevertheless changes their status from pairs to groups, due to the
presence of other dwarf companions. Note that the properties of multiple galaxies identified
using a certain criterion in the Local Volume differ from the corresponding properties of
multiple galaxies found in deep samples due to the decreasing detection rate of dwarf objects
with distance. That may be why the fraction of galaxies in binary systems, 12--17\%, in about
100~Mpc deep samples is somewhat higher than in the Local Supercluster  (10\%) or Local Volume
(7\%).

\section{CONCLUSIONS}

We identified the galaxy pairs that we included in our catalog without using the condition of
isolation. Therefore, as new dwarf galaxies are found in the volume of the Local Supercluster and
new line-of-sight velocity measurements are made for galaxies located in the vicinity of the pars,
the list of pairs will be updated by including new objects and some pairs will be promoted to
a higher multiplicity category. However, our sample still gives a correct idea about the kinematics of
the smallest, most simple systems within the 95-Mpc diameter volume.
About 40\% of the considered pairs have dimensionless parameters $II>5$, which allow us to treat these objects as
sufficiently isolated systems (Fig.~11). With the median mass of the pair,  $1.5\times
10^{11}M_{\odot}$, and median projected component separation of  $R_{\bot}=123$~kpc, the
typical density contrast in the ensemble of our pairs is  $\delta\rho/\rho_c\sim500$ in the units
of critical density. At such contrast systems of galaxies can be considered to be dynamically
detached from Hubble flow. On the other hand, with the median absolute value of the component
line-of-sight velocity difference  of 35 km/s and with the median projected separation of 123~kpc,
a typical pair of our catalog is characterized by the 
``crossing time'' of $3.5\times 10^9$~yr.
Hence the components of a typical pair could make about of four turns about the common mass center.

The use of the new clusterizetion algorithm, which takes into account individual
properties of galaxies, allowed us to discover a surprisingly large number of pairs
consisting of dwarf galaxies. Many components of these systems, which are located far from
normal galaxies, are rich in gas and are characterized by active star formation. Such
binary dwarf galaxies (for instance, IZw18) have been known long ago.
Among of them the galaxies with very low metallicity~\cite{Ekta06:Karachentsev_n} occur quite often.
Our list of galaxy triplets with velocities $V_{LG}<3500$ km/s also contains triple systems of dwarfs
with blue components. Many properties of our systems of dwarf galaxies do not differ from the
corresponding properties of associations of nearby dwarfs as described by
Tully et al.~\cite{Tal06:Karachentsev_n}. The closest example of the objects of this population is
located at the boundary of the Local group and includes NGC 3109, Sex A, Sex B, and Antlia. The
mutual velocities of these ``dark groups'' are equal to only 10 km/s, i.e., they are
comparable to velocity measurement errors. The evolutionary status of multiple dwarf
galaxies still remains totally unclear. According to the results of numerical simulations performed
by Bekki~\cite{Bek62:Karachentsev_n}, the evolution of such dwarf systems with extended gaseous
envelopes may be governed by their consequent mergers triggering the star-formation
bursts. We consider mass 21-cm line observations of these objects on aperture synthesis radio
telescopes with a resolution of about 1~km/s is very perspective.

\newpage
\input{Karachentsev_tab.tex}
\newpage

\twocolumngrid
\normalsize
\begin{acknowledgments}
This work was supported by the Russian Foundation for Basic Research (grant nos.~07--02--00005,
08--02--00627, and 07--02--00792) and  the joint grant (no.~06--02--04017) of Deutsche
Forschungsgemeinschaft and Russian Foundation for Basic Research. We acknowledge the use of the
HyperLeda database ({\tt http://leda.univ-lyon1.fr}) and NASA/IPAC Extragalactic Database (NED)
({\tt http://nedwww.ipac.caltech.edu/}).
We thank V.~E.~Karachentseva for finding the magnitudes and types for a large
number of dwarf galaxies.
\end{acknowledgments}

\end{document}

%% file: sao_cmd.tex
\def\saoname{Special Astrophysical Observatory of the Russian AS,
              Nizhnij Arkhyz 369167, Russia}

\def\squareforqed{\hbox{\rlap{$\sqcap$}$\sqcup$}}

\def\sq{\ifmmode\squareforqed\else{\unskip\nobreak\hfil
\penalty50\hskip1em\null\nobreak\hfil\squareforqed
\parfillskip=0pt\finalhyphendemerits=0\endgraf}\fi}

\def\sun{\hbox{$\odot$}}

\def\degr{\hbox{$^\circ$}}

\def\utw{\smash{\rlap{\lower5pt\hbox{$\sim$}}}}

\def\udtw{\smash{\rlap{\lower6pt\hbox{$\approx$}}}}

\def\diameter{{\ifmmode\mathchoice
{\ooalign{\hfil\hbox{$\displaystyle/$}\hfil\crcr
{\hbox{$\displaystyle\mathchar"20D$}}}}
{\ooalign{\hfil\hbox{$\textstyle/$}\hfil\crcr
{\hbox{$\textstyle\mathchar"20D$}}}}
{\ooalign{\hfil\hbox{$\scriptstyle/$}\hfil\crcr
{\hbox{$\scriptstyle\mathchar"20D$}}}}
{\ooalign{\hfil\hbox{$\scriptscriptstyle/$}\hfil\crcr
{\hbox{$\scriptscriptstyle\mathchar"20D$}}}}
\else{\ooalign{\hfil/\hfil\crcr\mathhexbox20D}}%
\fi}}

\newcommand{\aaa}{Astronom. and Astrophys.}
\newcommand{\aas}{Astronom. and Astrophys. Suppl. Ser.}

\newcommand{\aj}{Astronom. J.}
\newcommand{\apjs}{Astrophys. J. Suppl.}

\newcommand{\mnras}{Monthly Notices Roy. Astronom. Soc.}

\newcommand{\pasp}{Publ. Astronom. Soc. Pacific}
\newcommand{\pazh}{Pis'ma Astronom. Zh.}



%% file: Karachentsev_tab.tex
\clearpage
\onecolumngrid \begin{flushright} {\it APPENDIX} \\[-5pt] \end{flushright}
\renewcommand{\baselinestretch}{1}
\footnotesize

\begin{longtable*}{r|l|cc|r|r|r|r|r|r|r|r|r}
\caption{{\bf Table.} Catalog of binary galaxies in the Local Supercluster and its neighborhood.}\\

\hline
No.    &  Name                & J2000.0            & &$V_{LG}$    & $\pm$  & T & $K$ &  $R_\perp$ &$\log L$    & {$\frac{M}{L_K}$} & {$\frac{M^c}{L_K}$}  & $\log(II)$ \\
     &                               &                    & &\multicolumn{2}{c}{km s$^{-1}$} &   & mag & kpc        &{$L_\odot$}& {$\odot$} & {$\odot$} &           \\
\hline
\endfirsthead
\caption{{\bf Table.} (Contd.)}\\

\hline
No.    &  Name                & J2000.0            & &$V_{LG}$    & $\pm$  & T & $K$ &  $R_\perp$ &$\log L$    & {$\frac{M}{L_K}$} & {$\frac{M^c}{L_K}$}  & $\log(II)$ \\
     &                               &                    & &\multicolumn{2}{c}{km s$^{-1}$} &   & mag & kpc        &{$L_\odot$}& {$\odot$} & {$\odot$} &           \\

\hline
\endhead

\hline
\endfoot

\hline
\endlastfoot

   1 & NGC7820                       & J000430.8$+$051201 & & 3252 & 39 &  1 &  9.60 &   507 & 10.81 &   30.9 &  25.6 &     0.30 \\
     & UGC00027                      & J000428.8$+$055050 & & 3310 &  5 &  6 & 12.25 &       &       &        &       &          \\[3pt]
   2 & 6dF...                        & J000432.0$-$220503 & & 3154 & 74 & 10 & 13.05 &   217 &  9.55 &  320.8 &       &     1.41 \\
     & ESO538$-$021                  & J000545.9$-$220442 & & 3220 & 10 &  6 & 13.80 &       &       &        &       &          \\[3pt]
   3 & UGC00132                      & J001400.9$+$125746 & & 1896 &  9 &  8 & 12.85 &    72 &  9.08 &   15.1 &   5.7 &     0.78 \\
     & PGC138138                     & J001354.2$+$124827 & & 1910 &  5 & 10 & 14.70 &       &       &        &       &          \\[3pt]
   4 & ESO078$-$022                  & J002056.8$-$635126 & & 1632 & 20 &  4 & 11.18 &   226 &  9.64 &   37.5 &       &     0.88 \\
     & 6dF...                        & J001636.1$-$641138 & & 1608 & 74 &  9 & 12.50 &       &       &        &       &          \\
\end{longtable*}